\documentclass[a4,11pt]{article}
\usepackage{mathtext}
\usepackage[intlimits]{amsmath}
\usepackage[psamsfonts]{amssymb}
\usepackage[cp1251]{inputenc}
\usepackage[T2A]{fontenc}
\usepackage[english]{babel}
\usepackage{cite}
\usepackage{geometry}
\usepackage[a4,center]{crop}
\usepackage{enumerate}
\renewcommand{\d}{\mathrm{d}}
\newcommand{\e}{\mathrm{e}}
\renewcommand{\i}{\mathrm{i}}
\def\>{\rangle}
\def\<{\rangle}

\begin{document}

{\centering
{\Large \bf Random Solitons Realization of Quantum Mechanics and~Stochastic Qubits}
\bigskip

{\large  Yu.P. Rybakov$^1$, T.F. Kamalov$^2$}
\bigskip

{\small $^1$ Theoretical Physics Department People's Friendship University of Russia\\
   Mikluho-Maklay str., 6, Moscow, 117198, Russia\\
  e-mail: yrybakov@sci.pfu.edu.ru; yrybakov@mtu-net.ru}
\smallskip

{\small $^2$ Physics Department   Moscow State Opened University\\
     P. Korchagin str., 22, Moscow, 107996, Russia\\
  e-mail: ykamalov@rambler.ru}

}

\bigskip

\begin{abstract}
Stochastic realization of the wave function in quantum
mechanics, with the inclusion of soliton representation of extended particles, is
discussed. The concept of Stochastic Qubits is used for quantum computing modeling.
\end{abstract}

\section{Introduction. Wave---Particle Dualism and Solitons}

As a first motivation for introducing stochastic representation of the wave
function let us consider the de Broglie plane wave
\begin{equation}  \label{eq:1.1}
\psi = A\e^{-\i kx} = A\e^{-\i \omega t+\i(\mathbf{kr})}
\end{equation}
for a free particle with the energy $\omega$, momentum $\mathbf{k}$, and
mass $m$, when the relativistic relation
\begin{equation}  \label{eq:1.2}
k^2 = \omega^2-{\mathbf{k}}^2 = m^2
\end{equation}
holds (in natural units $\hbar = c = 1$).

Suppose, following L.~de~Broglie~\cite{1} and A.~Einstein~\cite{2}, that the
structure of the particle is described by a regular bounded function $u(t,
\mathbf{r})$, which is supposed to satisfy some nonlinear equation with the
Klein --- Gordon linear part. Let $l_0 = 1/m$ be the characteristic size of
the soliton solution $u(t, \mathbf{r})$ moving with the velocity $\mathbf{v}
= {\mathbf{k}}/{\omega}$.

Now it is worth-while to underline the remarkable fact behind this research~%
\cite{3}, namely, the possibility to represent the de Broglie wave (\ref%
{eq:1.1}) as the sum of solitons located at nodes of a cubic lattice with
the spacing $a \gg l_0$:
\begin{equation}  \label{eq:1.3}
A\e^{-\i kx} = \sum_{\mathbf{d}}u(t, \mathbf{r}+\mathbf{d}),
\end{equation}
where $\mathbf{d}$ marks the positions of lattice nodes. To show the
validity of (\ref{eq:1.3}) one can take into account the asymptotic behavior
of the soliton in its tail region:
\begin{equation}  \label{eq:1.4}
u(x) = \int\,\d^4 k \e^{-\i kx}g(k)\delta (k^2-m^2)
\end{equation}
and then use the well-known formula
\begin{equation}  \label{eq:1.5}
\sum_{\mathbf{d}}\e^{\i (\mathbf{k}\,\mathbf{d})} = {\left(\frac{2\pi}{a}%
\right)}^3 \delta (\mathbf{k}),
\end{equation}
implying that
\[
A = {\left(\frac{2\pi}{a}\right)}^3\frac{g(m)}{2m}.
\]
The formula (\ref{eq:1.3}) gives a simple illustration of the wave ---
particle dualism, showing that the de Broglie wave characterizes the
assemblage of particles --- solitons.

\section{Random Hilbert space}

The next step to get the stochastic representation of quantum mechanics can
be performed if one admits that the locations of solitons' centers are not
regular nodes of the cubic lattice but some randomly chosen points. To
realize this prescription, suppose that a field~$\phi${} describes $n$
particles --- solitons and has the form
\begin{equation}
\phi (t, \mathbf{r}) = \sum_{k = 1}^{n} \phi^{(k)} (t, \mathbf{r}),
\label{eq:2.1}
\end{equation}
where
\[
\text{supp}\, \phi^{(k)}\, \cap \,\,\text{supp}\, \phi^{(k^{\prime})} =
0,\quad k \ne k^{\prime},
\]
and the same for the conjugate momenta
\[
\pi (t, \mathbf{r}) = \partial \mathcal{L}/\partial \phi_t = \sum_{k =
1}^{n} \pi^{(k)} (t, \mathbf{r}), \quad \phi_t = \partial \phi/\partial t.
\]
Let us define the auxiliary functions
\begin{equation}
\varphi^{(k)} (t, \mathbf{r}) = \frac{1}{\sqrt{2}} (\nu_k \phi^{(k)} + i
\pi^{(k)}/\nu_k)  \label{eq:2.2}
\end{equation}
with the constants $\nu_k$ satisfying the normalization condition
\begin{equation}
\hbar = \int\, \d^3 x |\varphi^{(k)}|^2.  \label{eq:2.3}
\end{equation}
Now we define the analog of the wave function in the configurational space $%
\mathbb{R}^{3n} \ni \mathbf{x} = \{\mathbf{r}_1, \ldots \mathbf{r}_n\}$ as
\begin{equation}
\Psi_N (t, \mathbf{r}_1, \ldots \mathbf{r}_n) = (\hbar^n N)^{-1/2} \sum_{j =
1}^{N} \prod_{k = 1}^{n} \varphi_j^{(k)} (t, \mathbf{r}_k),  \label{eq:2.4}
\end{equation}
where $N \gg 1$ stands for the number of trials (observations) and $%
\varphi_j^{(k)}$ is the one-particle function (\ref{eq:2.2}) for the $j$ -
th trial. It can be shown~\cite{3,4} that the quantity
\[
\rho_N = \frac{1}{(\triangle \vee)^n}\, \int\limits_{(\triangle \vee)^n
\subset \mathbb{R}^{3n}}\d^{3n}x |\Psi_N|^2,
\]
where $\triangle \vee$ is the elementary volume which is supposed to be much
greater than the proper volume of the particle ${l_0}^3 = \vee_0 \ll
\triangle \vee$, plays the role of coordinate probability density.

If we choose the classical observable $A$ with the generator $\hat{M}_A$,
one can represent it in the form
\[
A_j = \int\, \d^3 x \pi_j \i \hat{M}_A \phi_j = \sum_{k = 1}^{n} \int\, \d^3
x \varphi_{j}^{* (k)} \hat{M}_{A}^{(k)} \varphi_{j}^{(k)},
\]
for the $j$ - th trial. The corresponding mean value is
\begin{multline}
\mathbb{E}(A) = \frac{1}{N} \sum_{j = 1}^{N} A_j = \frac{1}{N} \sum_{j =
1}^{N} \sum_{k = 1}^{n} \int\, \d^3 x \varphi_{j}^{* (k)} \hat{M}_{A}^{(k)}
\varphi_{j}^{(k)} = \int\,\d^{3n} x \Psi_{N}^{*} \hat{A} \Psi_{N} + O \left(\frac{\vee_0}{%
\triangle \vee}\right),  \label{eq:2.5}
\end{multline}
where the hermitian operator $\hat{A}$ reads
\begin{equation}
\hat{A} = \sum_{k = 1}^{n} \hbar \hat{M}_{A}^{(k)}.  \label{eq:2.6}
\end{equation}
Thus, upto the terms of the order $\vee_0/\triangle \vee \ll 1$, we obtain
the standard quantum mechanical rule (\ref{eq:2.5}) for the calculation of
mean values~\cite{4,5}.

It is interesting to underline that the solitonian scheme in question
contains also the well-known spin --- statistics correlation ~\cite{6}.
Namely, if $\varphi_{j}^{(k)}$ is transformed under the rotation by
irreducible representation $D^{(J)}$ of $SO (3)$, with the weight $J$, then
the transposition of two identical extended particles is equivalent to the
relative $2\pi$ rotation of $\varphi_{j}^{(k)}$, that gives the
multiplication factor $(-1)^{2J}$ in $\Psi_N$.

It can be also proved that $\Psi_N$ upto the terms of order $%
\vee_0/\triangle \vee$ satisfies the standard Schr\"{o}dinger equation ~\cite%
{6}. To verify the fact that solitons can really possess wave properties,
the \textit{gedanken} diffraction experiment with individual electrons ---
solitons was realized. Solitons with some velocity were dropped into a
rectilinear slit, cut in the impermeable screen, and the transverse momentum
was calculated which they gained while passing the slit, the width of which
significantly exceeded the size of the soliton. As a result, the picture of
distribution of the centers of scattered solitons was restored on the
registration screen, by considering their initial distribution to be uniform
over the transverse coordinate. It was clarified that though the center of
each soliton fell into a definite place of the registration screen
(depending on the initial soliton profile and the point of crossing the
plane of the slit by the center), the statistical picture in many ways was
similar to the well-known diffraction distribution in optics, i.e. Fresnel's
picture at short distances from the slit and Fraunhofer's one at large
distances~\cite{7,8}.

Fulfillment of the quantum mechanics correspondence principle for the
Einstein --- de Broglie's soliton model was discussed in the works \cite%
{4,5,6}.In these papers it was shown that in the framework of the soliton
model all quantum postulates were regained at the limit of point particles
so that from the physical fields one can build the amplitude of probability
and the average can be calculated as a scalar product in the Hilbert space
by introducing the corresponding quantum operators (\ref{eq:2.6}) for
observables. The fundamental role of the gravitational field in the
de~Broglie --- Einstein solitonian scheme was discussed in~\cite{6,9}. The
soliton model of the hydrogen atom was developped in~\cite{10,11}.

As a result we obtain the stochastic realization (\ref{eq:2.4}) of the wave
function $\Psi_N$ which can be considered as an element of the random
Hilbert space $\mathcal{H}_{\text{rand}}$ with the scalar product
\begin{equation}
(\psi_1, \psi_2) = \mathbb{E}(\psi_{1}^{*}\psi_2),  \label{eq:2.7}
\end{equation}
with $\mathbb{E}$ standing for the expectation value. As a rude
simplification one can admit that the averaging in (\ref{eq:2.7}) is taken
over random characteristics of particles --- solitons, such as their
positions, velocities, phases, and so on. It is important to underline once
more that the correspondence with the standard quantum mechanics is retained
only in the point -- particle limit ($\triangle \vee \gg \vee_0$) for $N \to
\infty$. To show this~\cite{4,5} one can apply the central limit theorem
stating that for $N \to \infty $\ $\Psi_N (t, \mathbf{x})$ behaves as the
Gaussian random field with the variance
\begin{equation}  \label{eq:2.8}
{\sigma }^2 = \rho (t, \mathbf{x}),\qquad \mathbf{x} \in \mathbb{R}^{3n},
\end{equation}
where $\rho (t, \mathbf{x})$ stands for the probability density (partition
function) of solitons' centers in $\mathbb{R}^{3n}$.

Random Hilbert spaces being widely exploited in mathematical statistics~\cite%
{12}, for quantum applications they were first used by N.~Wiener in~\cite%
{13,14}. To illustrate the line of Wiener's argument, we recall the general
scheme of introducing various representations in quantum mechanics.

Let $|\psi \>$~be a state vector in the Hilbert space $\mathcal{H}$ and $%
\hat{A}$~be a self-conjugate operator with the spectrum $\sigma (\hat{A})$.
Then the $a$ -- representation is given by the wave function
\[
\psi (a) = \<a|\psi \>,
\]
where
\[
\hat{A}|a\> = a|a\>,\qquad a \in \sigma (\hat{A}).
\]
In particular, the famous Schr\"{o}dinger coordinate $q$ -- representation
is given by the wave function
\begin{equation}  \label{eq:2.9}
\psi (q) = \<q|\psi \> = \sum_{n}\,\<q|n\>\<n|\psi \>,
\end{equation}
with $|n\rangle $ being some complete set of state vectors in $\mathcal{H}$.

Wiener considered the real Brownian process $x(s, \alpha )$ in the interval $%
[0, 1] \ni s$, where $\alpha \in [0, 1]$ is the generalized number of the
Brownian trajectory and the correlator reads
\begin{equation}  \label{eq:2.10}
\int_{0}^{1}\,\d \alpha \,x(s, \alpha )x(s^{\prime}, \alpha ) = \min \,(s,
s^{\prime}).
\end{equation}
To obtain the quantum mechanical description, Wiener defined the complex
Brownian process
\begin{equation}  \label{eq:2.11}
z(s|\alpha , \beta ) = \frac{1}{\sqrt{2}}\left[x(s, \alpha )+\i\,y(s, \beta )%
\right]; \alpha ,\,\beta \in [0, 1],
\end{equation}
and using the natural mapping $\mathbb{R}^3 \to [0, 1]$, for the particle in
$\mathbb{R}^3$, constructed the stochastic representation of the wave
function along similar lines as in (\ref{eq:2.9}):
\begin{equation}  \label{eq:2.12}
\langle \alpha , \beta |\psi \rangle = \int_{s \in [0, 1]}\,\d z(s|\alpha , \beta )\psi
(s),
\end{equation}
with the obvious unitarity property
\[
\int_{0}^{1}\,\d s\,|\psi (s)|^2 = \iint\limits_{[0,
1]^2}\,\d\alpha\,\d\beta |\<\alpha ,\beta |\psi \>|^2
\]
stemming from (\ref{eq:2.10}).

\section{Stochastic qubits and solitons}

Now we intend to explain how stochastic qubits (quantum bits) could be
introduced in the solitonian scheme. To this end one should define the
random phase $\Phi_j$ for the $j$ - th trial in our system of $n$ solitons
--- particles. Let ${\varphi }^{(k)}(\mathbf{r})$ denote the standard
(etalon) profile for the $k$ - th soliton. The most probable position $%
\mathbf{d}_j^{(k)}(t)$ of the $k$ - th soliton's center in $j$ - th trial
can be found from the following variational problem:
\begin{equation}  \label{eq:3.1}
\left|\int\,\d^3 x\,{\varphi }_j^{*(k)}(t, \mathbf{r}) {\varphi }^{(k)}\left(%
\mathbf{r}-\mathbf{d}_j^{(k)}\right)\right|\quad\to\quad \max,
\end{equation}
thus giving the random phase structure:
\begin{equation}  \label{eq:3.2}
\Phi_j = \sum_{k = 1}^{n}\,\text{arg}\,\int\,\d^3 x\,{\varphi }_j^{*(k)}(t,
\mathbf{r}) {\varphi }^{(k)}\left(\mathbf{r}-\mathbf{d}_j^{(k)}\right).
\end{equation}

The random phase (\ref{eq:3.2}) can be used for simulating quantum computing
via generating the following $M$ random dichotomic functions:
\begin{equation}  \label{eq:3.3}
f_s(\theta_s) = \text{sign}\,\left[\cos (\Phi_j+\theta_s)\right],\qquad s =
\overline{1, M},
\end{equation}
with $\theta_s$ being arbitrary fixed phases. Now recall that the quantum
bit (qubit) is identified with the state vector
\[
|\psi \> = \alpha |0\> + \beta |1\>,
\]
corresponding to the superposition of two orthogonal states $|0\>$ and $|1\>$,
as for instance, two polarizations of the photon, or two possible $1/2$ -
spin states. Using the well-known quantum expression for $1/2$ - spin
correlation in a singlet state of two particles (the latter being a typical
entanglement state):
\begin{equation}  \label{eq:3.4}
\mathbb{E}\,(\sigma_a \,\sigma_b) = -(\mathbf{a\,b}),\qquad |\mathbf{a}| = |%
\mathbf{b}| = 1,
\end{equation}
one can compare it with the random phases correlation for the case of $n = 2$
particles~\cite{15}:
\begin{equation}  \label{eq:3.5}
\mathbb{E}\,(f_1 f_2) = 1 - \frac{2}{\pi}|\triangle \theta |,
\end{equation}
where $\triangle \theta = \theta_1 - \theta_2$. The similarity of these two
functions (\ref{eq:3.4}) and (\ref{eq:3.5}) of the angular variable seems to
be a good motivation for the $M$ qubits simulation by the dichotomic random
functions (\ref{eq:3.3}), popularized in the paper~\cite{15}.

In conclusion we express the hope that the random solitons realization of
the wave function could be effectively used for quantum computing simulation.

\end{document}